# Measuring autoionization decay lifetimes of optically forbidden inner valence excited states in neon atoms with attosecond noncollinear four wave mixing spectroscopy

Nicolette G. Puskar[1,2], Yen-Cheng Lin[1,2], James D. Gaynor[1,2], Maximilian C. Schuchter[3], Siddhartha Chattopadhyay[4], Carlos Marante[4], Ashley P. Fidler[5], Clare L. Keenan[6], Luca Argenti[4,7], Daniel M. Neumark[1,2,*] and Stephen R. Leone[1,2,8,*]

[1]Department of Chemistry, University of California, Berkeley, California 94720, USA
[2]Chemical Sciences Division, Lawrence Berkeley National Laboratory, Berkeley, California 94720, USA
[3]Laboratory for Physical Chemistry, ETH Zürich, Zürich 8093, Switzerland
[4]Department of Physics, University of Central Florida, Orlando 32816, USA
[5]Department of Chemistry, Princeton University, New Jersey 08544, USA
[6]Department of Chemistry, The University of Chicago, Illinois 60637, USA
[7]CREOL, University of Central Florida, Orlando 32816, USA
[8]Department of Physics, University of California, Berkeley, California 94720, USA
*Authors to whom correspondences may be addressed: (D.M.N.) dneumark@berkeley.edu; (S.R.L) srl@berkeley.edu

**Abstract** Attosecond noncollinear four wave mixing spectroscopy with one attosecond extreme ultraviolet (XUV) pulse and two few-cycle near-infrared (NIR) pulses was used to measure the autoionization decay lifetimes of inner valence electronic excitations in neon atoms. After a 43-48 eV XUV photon excites a 2$s$ electron into the 2$s$2$p^6$[n$p$] Rydberg series, broadband NIR pulses couple the 2$s$2$p^6$3$p$ XUV-bright state to neighboring 2$s$2$p^6$3$s$ and 2$s$2$p^6$3$d$ XUV-dark states. Controllable delays of one or both NIR pulses with respect to the attosecond XUV pulse reveal the temporal evolution of either the dark or bright states, respectively. Experimental lifetimes for the 3$s$, 3$p$, and 3$d$ states are measured to be 7 ± 2 fs, 48 ± 8 fs, and 427 ± 40 fs, respectively, with 95% confidence. Accompanying calculations with two independent *ab initio* theoretical methods, NewStock and ASTRA, verify the findings. The results support the expected trend that the autoionization lifetime should be longer for states that have a smaller penetration in the radial region of the 2$s$ core hole, which in this case is for the higher angular momentum Rydberg orbitals. The underlying theory thus links the lifetime results to electron correlation and provides an assessment of the direct and exchange terms in the autoionization process.

**Introduction**

Autoionization has played a crucial role in the development of atomic theory; [1] the process is the spontaneous emission of an electron by a system excited to a discrete configuration with energy above the ionization threshold, due to the interaction with configurations in the electronic continuum [2]. The lifetime of an autoionizing state can be described quantum mechanically using time-dependent perturbation theory. In atoms, which feature only electronic degrees of freedom, the relevant matrix element describes autoionization as a process dependent on the interaction of two electrons where the perturbation is Coulombic repulsion [3]. According to Fermi's golden rule [4], the autoionization rate is given as

$$\Gamma = 2\pi \left| \langle \psi_f(r_1)\phi_f(r_2) \left| \frac{1}{|r_1 - r_2|} \right| \psi_i(r_1)\phi_i(r_2) \rangle \right|^2 \tag{1}$$

where $\psi_i(r_1)$ and $\phi_i(r_2)$ are the initial states of the two electrons, $\psi_f(r_1)$ and $\phi_f(r_2)$ are the final bound and continuum states of the electrons, and $\frac{1}{|r_1-r_2|}$ is the electrostatic repulsion between the two electrons. [5] Hence, atomic autoionization is a fundamental probe of electron-electron correlation.

The lifetimes of an autoionizing states are also explored experimentally; the interference between the discrete and continuum states forms asymmetric resonance peaks known as Fano resonances. [3,6] The resonance width (or linewidth) of atoms is a descriptive parameter that can be used to determine the decay lifetime of the discrete state via the time-energy uncertainty principle. Thus, frequency-domain linewidth measurements in atoms have been widely used to infer ultrafast autoionization decay lifetimes. [7]

Time-resolved autoionization dynamics is now experimentally feasible with sub- and few-femtosecond temporal resolution using attosecond pulses with extreme ultraviolet (XUV) photon energies. [8–11] While this development has deepened our understanding of dynamics in optically allowed (bright) states, the study of autoionization dynamics in optically forbidden (dark) states has been much less frequently studied due to the optical selection rules for one photon absorption that govern the excitation. In this work, we utilize the ability of attosecond noncollinear four wave mixing (FWM) spectroscopy, which involves multiple photon excitation processes, to directly measure time-resolved autoionization decay lifetimes in both bright and dark inner valence-excited states. This work benchmarks lifetime dynamics in neon atoms by accurate comparison to theory.

The Ne $2s2p^6[np]$ autoionizing Rydberg series was first studied experimentally with synchrotron radiation by Codling [12] *et al* and was later repeated with higher resolution by Schulz [13] *et al*. Theoretical studies of this Rydberg series with the R-matrix method [14] and density functional theory [15] characterized the resonance linewidths for the $2s2p^63p$ state, providing a decay lifetime in the range of 41-55 fs. In contrast, very few studies of the $2s2p^63s$ and $2s2p^63d$ states are available. There is one linewidth measurement of the dark $2s2p^63s$ state using electron energy loss spectroscopy (Min [16] et al, corresponding lifetime of 7 fs), and no existing measurement of the dark $2s2p^63d$ state is known. It is worth noting that Ding [17] et al studied the coupling dynamics of these same states with experiment and theory, but did not report autoionizing decay lifetimes.

Attosecond FWM spectroscopy utilizes a noncollinear beam geometry between a generated attosecond XUV pulse and two few-cycle NIR pulses to yield background-free, spatially isolated emission signals [18] (Figure 1a). Atomic studies using attosecond FWM spectroscopy have obtained results on the evolution of coherences between odd ($1snp$) and even ($1sns/1snd$) parity states in helium, measured the lifetimes of multiple $^2P_{1/2}$ $ns/d$ autoionizing states in krypton, and explored non-resonant coherent amplitude transfer in argon. [19–21] Molecular studies using FWM spectroscopy have investigated the dipole-forbidden double-well potential of the a" $^1\Sigma_g^+$ dark state of $N_2$, extracted few femtosecond lifetimes for the v=0 and v=1 $3s$ Rydberg states in $O_2$, and probed the dynamics of autoionizing inner valence excited Rydberg states in $CO_2$. [22–24] These previous FWM experiments have successfully studied dynamics excited in the 11 eV - 29 eV range. However, an array of atomic and molecular dynamics involving inner valence and core-excited states are accessible at higher XUV energies. Here, we investigate the $2s2p^6[np]$ Rydberg series in neon that is excited using photons of 43-48 eV.

The neon $2s2p^6[np]$ autoionizing Rydberg series is an excellent system to probe dark state decay lifetimes due to the close proximity of the $3s$ and $3d$ dark states to the $3p$ bright state, within the bandwidth of the NIR pulses used in the FWM experiment. After a $2s$ electron is excited into the $3p$ bright state with an XUV pulse, the $3s$ or $3d$ dark states can be resonantly accessed with broadband NIR photons (~1.3 – 2.0 eV) in a V- or Λ-coupling scheme, respectively (Figure 1b). The $2s^{-1}2p^6$ inner valence electronic configuration accessed in this series of FWM experiments allows us to relate autoionization decay lifetimes to a series of orbital angular momentum states that have differing penetration in the core hole $2s$ state radial region, while the principal quantum numbers are held constant. In other words, the flexibility of controlling each light-matter interaction in the FWM pulse sequence enables the systematic investigation of autoionization as a function of the orbital geometry (angular momentum states l = 0, 1, and 2) using FWM photon-emission detection.

This study reports time-resolved autoionization decay lifetime measurements of optically bright and dark states using attosecond four wave mixing spectroscopy. The lifetimes for the $2s^{-1}2p^6$ $3s$, $3p$, and $3d$ states of Ne are determined to be $7 \pm 2$ fs, $48 \pm 8$ fs, and $427 \pm 40$ fs respectively with a 95% confidence interval. *Ab initio* theoretical calculations with the atomic code NewStock and molecular ionization code ASTRA are also performed, which verify the experimental findings. Furthermore, an analytical hydrogenic model is used to estimate the separate contribution of the direct and exchange terms to the configuration-interaction matrix element responsible for the autoionization. The results depict a direct relationship between orbital angular momentum and autoionization decay lifetime based on orbital penetration: as angular momentum increases, the overlap with the $2s$ core hole decreases, and the autoionization lifetimes are therefore expected to increase [25,26]. The lifetimes reported in this work support this trend. As exemplified here, attosecond FWM can provide time-domain insight into highly excited electronic dynamics in both bright and dark states with few-femtosecond temporal resolution in a background-free manner, facilitating a greater physical understanding of ultrafast relaxation processes.

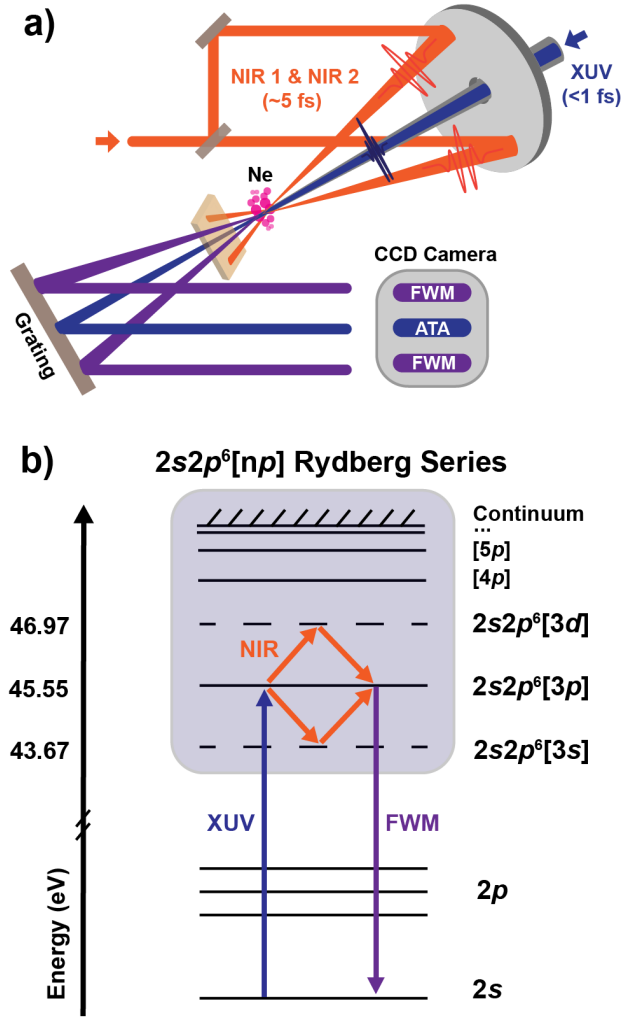

*Figure 1.* Experimental Schematic. (a) The attosecond XUV + NIR pulse interactions leading to four wave mixing generation. A noncollinear beam geometry results in spatially distinct FWM emission signals. (b) An energy level diagram of the neon $2s2p^6[np]$ Rydberg series, which lies above the first ionization limit at 21.56 eV [27]. The solid lines represent bright states and the dotted lines represent dark states. The dark states are accessible by resonant coupling with the broadband NIR pulses.

**Experimental Methods**

A 13 mJ/pulse, 1 kHz repetition rate Ti:Sapphire commercial laser system (Legend Elite Duo HE, Coherent) produces 40 fs NIR pulses at a central wavelength of 800 nm. The output pulses are spectrally broadened via self-phase modulation in a 700 μm inner diameter, stretched hollow core fiber filled with 934 Pa of neon gas. The pulses are then recompressed by eight pairs of double-angled chirped mirrors (PC70, Ultrafast Innovations), producing broadband NIR pulses spanning 550-950 nm with < 5 fs pulse duration. A 70:30 beam splitter divides the broadband light into two beams. The reflected light (70%) is focused with a spherical mirror (f = 50 cm) into a vacuum chamber at $1.34 \times 10^{-5}$ Pa that contains a cell with argon gas for high harmonic generation (HHG). After HHG generates attosecond XUV pulses, an Al foil (0.20 μm thick, Lebow) filters out the co-propagating NIR beam, and the XUV pulses are focused via a gold-coated toroidal mirror (f = 50 cm) into the sample cell filled with neon gas (1,467 Pa backing pressure).

The NIR pulse transmitted by the beam splitter (30%) is delayed relative to the XUV with a piezoelectric stage and is then split again by a 50:50 beam splitter. After one of these beams is further delayed by a second piezoelectric stage, two spherical mirrors (f = 1 m) direct the beams independently toward the in-vacuum recombination mirrors to spatially overlap the two focused NIR pulses at the sample cell interaction region. To ensure spatial separation of the FWM signal, the two NIR beams intersect the XUV light in a noncollinear geometry at the same angles of approximately 22.7 mrad (1.3°). After the FWM signal is generated, the NIR beams are attenuated with a second Al filter (0.15 μm thick, Lebow) while the transmitted XUV light and wave-mixing signals are frequency-dispersed by a gold-coated flat-field variable line space grating. The spectral signals are then recorded as a function of frequency and phase-matching divergence angle by a 1340 x 400-pixel X-ray CCD camera (Pixis XO 400B, Princeton Instruments).

Temporal and spatial overlap between the XUV and two NIR beams is determined using second harmonic generation signals with an in-vacuum BBO crystal in the sample chamber. Given that this is a three-pulse experiment, there are two time delays available to manipulate: the first time delay ($\tau_1$) is between the XUV and one NIR (NIR1) beam, while the second time delay ($\tau_2$) is between NIR1 and the second NIR (NIR2) beam. Negative time delay indicates that the NIR beams arrive before the XUV pulse, and positive time delay indicates that NIR beams arrive after the XUV pulse. Precise control of these different time delays exploits the phase matching condition (Equation 11), which governs whether the bright or dark state lifetime is contained in the FWM signal (Figure 3).

To measure the lifetime of the 3s state, delays between [-200 fs to 50 fs] were scanned in 1 fs steps. For the 3p state, delays between [-60 fs – 150 fs] were scanned in 1 fs steps. For the 3d state, delays between [-250 fs – 1100 fs] were scanned in 20 fs steps at long negative time delays, 2 fs steps during the rise and initial decay, and 10 fs steps during the long positive time delays. In order to maintain as similar conditions as possible during scan collection, the neon sample pressure was maintained at 1,334 – 1,600 Pa and the average NIR pulse power was kept in the range possible of 0.7-1.0 mW (1 kHz repetition rate) to minimize any strong-field effects that could impact the results.

**Theoretical Methods**

To compute the lifetime of the autoionizing states for inner shell ionization from Ne, we use two complementary *ab initio* close-coupling approaches: the NewStock suite of atomic photoionization codes [28], in association with the exterior complex scaling (ECS) technique [29–31], and the new ASTRA [32] molecular ionization code, in association with complex absorption potentials (CAPs). The predictions of these two codes, which employ different basis sets and methods, validate each other. Furthermore, to rationalize the trends observed as a function of the angular momentum both in experiment and in *ab initio* theory, we estimate the width of the relevant resonances with an analytical hydrogenic model that is able to discriminate direct and exchange contributions. In the following, we offer a brief description of these three approaches.

**NewStock Calculations:** NewStock represents single-ionization states using a close-coupling (CC) expansion with pseudostates, obtained from the anti-symmetrized product of a finite set of parent ionic states with a single-particle states,

$$|\Psi(x_1,-,x_N)\rangle = \hat{\mathcal{A}} \sum_{\Gamma\alpha} \Phi_\alpha^\Gamma(x_1,-,x_{N-1};\widehat{r_N},\zeta_N)\varphi_\alpha^\Gamma(r_N) + \sum_{\Gamma i} \chi_i^\Gamma(x_1,-,x_N)c_i^\Gamma \quad (2)$$

where $\hat{\mathcal{A}}$ is the antisymmetrizer, $x_i = (\vec{r}_i, \zeta_i)$ are the spatial and spin coordinates of the $i$-th electron, and $r = \sqrt{\vec{r}\cdot\vec{r}}$, $\hat{r} = \vec{r}/r$. The channel functions $\Phi_\alpha^\Gamma$ are obtained by coupling the orbital and spin coordinates of the ion and the photoelectron to a well-defined orbital angular momentum $L$ and spin $S$.

The collective index $\Gamma$ represents the overall state symmetry, i.e., parity, the total spin and orbital angular momentum and their projections. The channel index $\alpha$ identifies, within $\Gamma$, a specific ionic state and photoelectron angular momentum $l_\alpha$. Equation [2] includes a last term formed by N-electron localized configurations, $\chi_i(x_1, -, x_N)$, which complete the description of the many-body function close to the nucleus. The active ionic orbitals are computed using the multi-configuration Hartree-Fock method (MCHF) with the ATSP2K package [33]. NewStock expresses the radial part of all bound and continuum orbitals in ECS B-splines [34], with complex-rotation angle $\theta$ and a scaling radius $R_0$ chosen such that all MCHF orbitals are negligible for $r > R_0$. The position $E_n$ and width $\Gamma_n$ of the resonances of interest are obtained from the real and imaginary part of the complex eigenvalues of the Hamiltonian in the ECS basis, $\widetilde{E_n} = E_n - i\frac{\Gamma_n}{2}$.

The CC channels are built by coupling the first four lowest states of the Ne$^+$ ion, with dominant configuration $2s^2 2p^5 (^2P^o)$, $2s 2p^6 (^2S^e)$, $2s^2 2p^4 3s (^2P^e)$, and $2s^2 2p^4 3s (^2D^e)$, with photoelectron states with $l_\alpha < 4$. To optimize the parent-ion states, we employ the MCHF method in a configuration space that includes all possible single and double excitations from the $2s^2 2p^5$ reference determinant to orbitals with $n \leq 3$. The energies thus obtained for the $2s^2 2p^5 (^2P^o)$ and $2s 2p^6 (^2S^e)$ states differ only by 8 meV from the NIST values [35]. We use a quantization box with radius $R_{\text{box}} = 300$ arb. units, an ECS radius $R_0 = 80$ arb. units, and an ECS angle $\theta = 0.1$ rad. To test if the calculations are reasonably converged, we have extended the calculations for the $^2S^e$ symmetry, which is the least numerically expensive, to include single and double excitations to $n = 4$, or triple excitation to $n = 3$, finding a variation of 10% or less on the lifetime of the $2s 2p^6 3s$ resonance, which is within the experimental error.

**ASTRA Calculations:** To confirm NewStock predictions for the resonances lifetimes, we used ASTRA [32], an independent theoretical suite of *ab initio* CC codes, based on a new Transition Density Matrices (TDMs) formalism, which can describe both atomic and molecular ionization. While NewStock takes into account SO(3) symmetry to the full, ASTRA is restricted to the D$_{2h}$ abelian point group, as it is a common practice in quantum chemistry codes [36–38].

In ASTRA, CC channels are obtained as spin-coupled antisymmetrized products of a molecular ionic state $|A\rangle$, with $N-1$ electrons, and a spin-orbital $|p\,\pi\rangle$ for an additional $N$-th electron in the orbital $p$ and with spin projection $\pi$ along the quantization axis, that can either be bound or lie in the continuum:

$$|A,p;S\Sigma\rangle = \sum_{\Sigma_A \pi} C^{S\Sigma}_{S_A \Sigma_A, \frac{1}{2}\pi} a^\dagger_{p\pi} |A_{\Sigma_A}\rangle \qquad (3)$$

where $C^{S\Sigma}_{S_A \Sigma_A, \frac{1}{2}\pi}$ are Clebsch-Gordan coefficients and $a^\dagger_{p\pi}$ is the creator operator [39] for an electron in orbital $p$ with spin projection $\pi$. The correlated ionic states $|A_{\Sigma_A}\rangle$ are computed at the Complete Active Space Configuration-Interaction (CASCI) level with LUCIA, a large-scale general CI code [40–43]. Here, the ionic orbitals are obtained with a Hartree Fock calculation of the Ne ground state, in a cc-pVTZ basis, using the DALTON program [36,37]. The remaining orbitals, which are needed to describe Rydberg satellites and the photoelectron, are expanded in a hybrid Gaussian-B-spline basis, with angular momentum up to $\ell_{max} = 3$, in a 400 arb. units quantization box [44]. The one- and two-electron integrals in the hybrid basis are computed using the GBTO library [45]. The active-orbital sector of general operators in the CC basis are computed with LUCIA, whereas ASTRA evaluates the remaining matrix elements from the basic electronic integrals and from the first- and second-order TDMs between correlated ions, which LUCIA computes with high efficiency.

To determine the position and width of autoionizing states, we enforce Siegert boundary conditions [46,47] on the eigenstates of the Hamiltonian by adding to it a complex absorption potential $V_{CAP}$, $\widetilde{H} = H + V_{CAP}$, defined as

$$V_{CAP} = -ic \sum_i \theta(r_i - R_{CAP})(r_i - R_{CAP})^2 \quad c \in \mathbb{R}^+_0 \qquad (4)$$

where $\theta(x)$ is the Heaviside step function. Convergence studies show that, with a suitable choice of the parameters $R_{CAP}$ and c, the complex absorber reproduces the ECS results [48]. As for the NewStock calculations, the CC basis used for neon in the ASTRA calculation includes the first four parent ions, with $^2P^o$, $^2S^e$, $^2P^e$ and $^2D^e$ symmetry. The active space comprised all the orbitals up to $n = 3$ principal quantum number, keeping the $1s$ doubly occupied.

**Hydrogenic model:** The width of the $2s^{-1}3L$ autoionizing states ($L = s, p, d$) is given by their interaction with the continuum channels into which they decay [Eq. (1)],

$$\Gamma_{2s^{-1}3L} = 2\pi \sum_{\ell=L\pm 1} \left|\left\langle \Psi_{2p^{-1}\varepsilon\ell} \left| \frac{1}{|r_1 - r_2|} \right| \Psi_{2s^{-1}3L} \right\rangle\right|^2 \quad (5)$$

where $2p^{-1}\varepsilon\ell$ represents the dominant configuration of a continuum channel defined by the $^2P^o$ ionic state coupled to a photoelectron with angular momentum $\ell$, with photoelectron energy $\varepsilon$ chosen so that the (antisymmetrized) continuum configuration $2p^{-1}\varepsilon\ell$ and the discrete state with dominant configuration $2s^{-1}3L$ are degenerate.

If we approximate both the discrete and the continuum states in Eq. [5] with their dominant configuration, the resonance width can be rewritten as the sum of a direct term $J_\ell$ and an exchange term $K_\ell$ between unsymmetrized configurations,

$$\Gamma_{2s^{-1}3L} \approx 2\pi \sum_{\ell=L\pm 1} |J_\ell - K_\ell|^2 \quad (6)$$

where

$$\begin{aligned} J_\ell = \langle 2s\varepsilon\ell|r_{12}^{-1}|2p3L\rangle &= \int d^3r_1 d^3r_2 \frac{[2s(\vec{r}_1)2p(\vec{r}_1)] \times [\varepsilon\ell(\vec{r}_2)3L(\vec{r}_2)]}{|\vec{r}_2 - \vec{r}_1|} \\ K_\ell = \langle 2s\varepsilon\ell|r_{12}^{-1}|3L2p\rangle &= \int d^3r_1 d^3r_2 \frac{[2s(\vec{r}_1)3L(\vec{r}_1)] \times [\varepsilon\ell(\vec{r}_2)2p(\vec{r}_2)]}{|\vec{r}_2 - \vec{r}_1|} \end{aligned} \quad (7)$$

Both the direct and exchange term, therefore, represent the Coulomb repulsion between two different charge densities. In the direct term, for example, these charge densities are the product of two $n = 2$ orbitals and that of the $3L$ and a continuum orbital. The interaction between these charge densities is inversely proportional to their separation. In particular, while only the exchange term contains an explicit product of the orbital of the electron satellite in the autoionizing state and of the 2s hole that is filled by the decay, both terms are enhanced by the 3L orbital penetrating the radial region where the n = 2 shell orbitals are located. This approximation allows us to estimate the contribution of the direct and exchange terms to the resonance widths $\Gamma_{2s^{-1}3L}$, which exhibit a marked decrease as L increases. To do so, we use hydrogen-like wave functions to model both the bound $\psi_{n\ell m}(\vec{r}) = u_{n\ell}(r)Y_{\ell m}(\hat{r})/r$ and the continuum orbitals $\psi_{E\ell m}(\vec{r}) = u_{E\ell}(r)Y_{\ell m}(\hat{r})/r$. For the bound orbitals [49],

$$u_{n\ell}(r) = N_{n\ell}\rho^{\ell+1}e^{-i\rho}\ _1F_1(\ell + 1 - i\eta; 2\ell + 2; 2i\rho)$$

$$N_{n\ell} = \frac{2^{\ell+1}}{ni^{\ell+1}(2\ell + 1)!}\sqrt{\frac{Z_{eff}(n+\ell)!}{(n-\ell-1)!}}$$

(8)

where $_1F_1(\alpha; \beta; z)$ is the confluent hypergeometric function [50], $\rho = \kappa r$ with $\kappa = iZ_{eff}/n$ and $\eta = in$. The effective charge $Z_{eff}$ is set to 5.7584 for $n = 2$ [51,52] and to 1.0 for n > 2. For the continuum orbitals,

$$u_{E\ell}(r) = C_{E\ell}\rho^{\ell+1}e^{-i\rho}\ _1F_1(\ell + 1 - i\gamma; 2\ell + 2; 2i\rho)$$

$$C_{E\ell} = 2^\ell e^{-\frac{\pi}{2}\gamma}\sqrt{\frac{2}{\pi k}}\frac{|\Gamma(\ell + 1 - i\gamma)|}{(2\ell + 1)!}$$

(9)

where $\rho = kr$ with $k = \sqrt{2E}$ and $\gamma = -k^{-1}$. Using these hydrogenic functions we obtained for the $2s^{-1}3s$, $2s^{-1}3p$ and $2s^{-1}3d$ autoionizing states the following lifetimes: 5 fs, 47 fs and 756 fs.

**Results and Discussion**

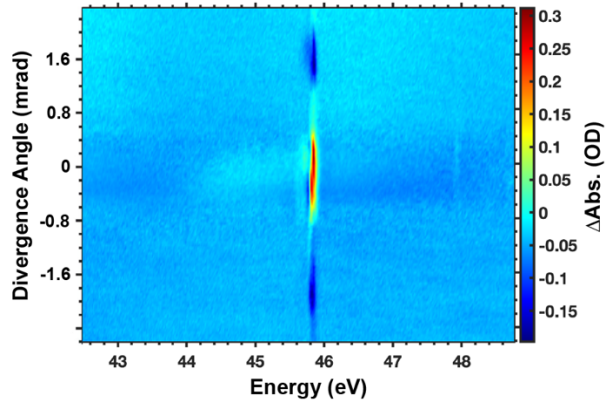

*Figure 2.* Camera image of the FWM signal from the Ne $2s2p^63p$ state at 45.55 eV at temporal overlap ($\tau_1 = \tau_2 = 0$). The red feature indicates transient absorption signals while the blue features indicate FWM emission signals.

The FWM signal of the Ne $2s2p^63p$ state at spatial and temporal overlap of all pulses is presented in Figure 2. To generate this camera image, the light is spectrally dispersed on the x-axis by the XUV grating and vertically separated on the y-axis by divergence angle due to the phase-matching process produced by the noncollinear beam geometry. The image is quantified in terms of differential absorption

$$\Delta A = -log_{10}\left(\frac{I}{I_0}\right) \quad (10)$$

where $I$ is the signal intensity with NIR and XUV beams incident on the sample and $I_0$ is the reference signal intensity when the two NIR beams are blocked by an automated shutter. So, the $I_0$ spectrum is composed of the dark counts from the CCD and the XUV pulse only. Absorption is indicated by the positive features (red) while emission is indicated by the negative features (dark blue). The four wave mixing signals are all in emission.

    The camera image shows three main signals appearing at different divergence angles. The attosecond transient absorption signal (red) of neon appears collinear with the XUV pulse wavevector, which is set to be 0 mrad. The upper and lower signals of negative intensity at ±1.6 mrad are FWM signals (dark blue). The energy and angle of the FWM signals are a result of the

conservation of energy and momentum, respectively, of all incident photons, which gives the phase matching condition for V- type and Λ-type couplings, respectively:

$$k_{FWM} = k_{XUV} \pm k_{NIR1} \mp k_{NIR2} \qquad (11)$$

The divergence angle of these FWM signals can be calculated using the expression:

$$\phi_{div,\Lambda} \approx \frac{E_{NIR1}\theta_{NIR1} + E_{NIR2}\theta_{NIR2}}{E_{XUV}} \qquad (12)$$

where $E_{NIR1}$ and $E_{NIR2}$ are the photon energies of the two NIR pulses, $E_{XUV}$ is the photon energy of the XUV pulse, and $\theta_{NIR1}$ and $\theta_{NIR2}$ are the crossing angles of the two NIR pulses with the XUV pulse. It is worth noting that in Eq. 12, the approximate linear dependency of the FWM divergence angle on the NIR photon energy is valid due to the small $E_{NIR}/E_{XUV}$ ratio.

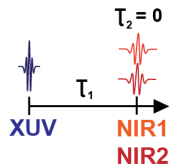
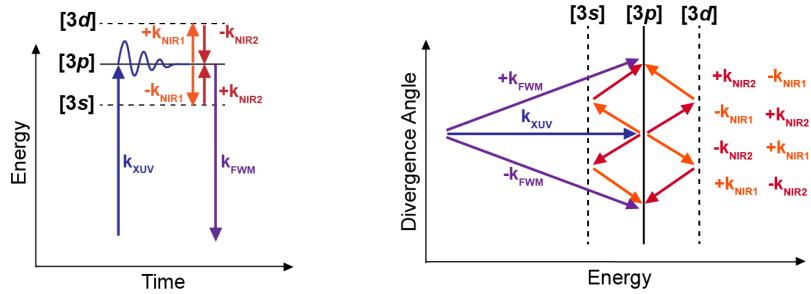
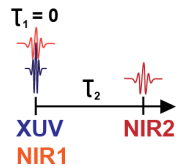
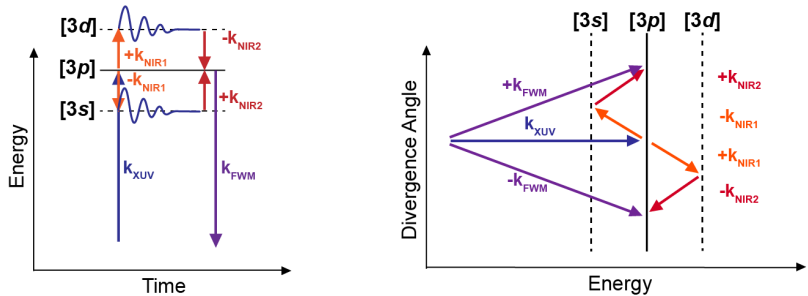

*Figure 3.* Pulse sequences for (a) bright and (b) dark state measurements. The left of the image shows the different delay choices between each of the three pulses. The middle of the image shows how we can select the state lifetime desired depending on the pulse sequence chosen. The right of the image shows how the phase matching condition determines the divergence angle of the emitted FWM signal. The solid lines represent the XUV-bright [3p] state while the dashed lines represent the XUV-dark [3s] and [3d] states.

Different pulse sequences are used to measure either the bright state or the dark state dynamics (Figure 3). [53] In a bright state scan (Figure 3a), the XUV pulse excites the sample and the two NIR beams are delayed simultaneously, but held at temporal overlap with each other; this approach is used to reveal the dynamics of the bright 3p state (XUV-NIR1 delay labeled $\tau_1$). In a dark state scan (Figure 3b), the XUV pulse and NIR1 beam are held at temporal overlap to excite an electron first to the 3p state and then immediately couple it to either the 3s or the 3d dark states. The dark state dynamics are scanned by delaying the second NIR beam relative to the first NIR beam (NIR1-NIR2 delay labeled $\tau_2$).

The FWM signals are generated through the V- and Λ-coupling pathways shown explicitly in Figure 3. First, XUV pulses excite an electron from the inner valence 2s state into the $2s2p^63p$ XUV-bright state. Then, NIR photons couple the 3p XUV-bright state to the 3s and 3d XUV-dark states. Finally, delaying the NIR pulses in either the bright or dark state pulse sequence reveals the transient behavior of the selected state. The FWM emission is then observed at 45.55 eV (the 3p state) as the system completes the coupling pathway and returns to the ground state.

The phase matching condition is responsible for why we can obtain independent dark state lifetime information from the positive and negative FWM signals. [53] On axis, many FWM signals are overlapped with attosecond transient absorption, making it difficult to resolve the FWM signals alone. In a bright state scan ($\tau_1$ is delayed, $\tau_2 = 0$), FWM signals at both positive and negative angles, above and below the on-axis signal, contain equivalent information for the 3p state. However, in a dark state scan ($\tau_1 = 0$, $\tau_2$ is delayed), this information symmetry is broken. Because the 3s and 3d dark states exist above and below the 3p bright state, the phase matching condition provides state-specific lifetime measurements for the positive and negative FWM signals. In this, the 3s state is probed through the upper wave mixing signal (+ 1.6 mrad) and the 3d state is probed through the lower wave mixing signal (- 1.6 mrad).

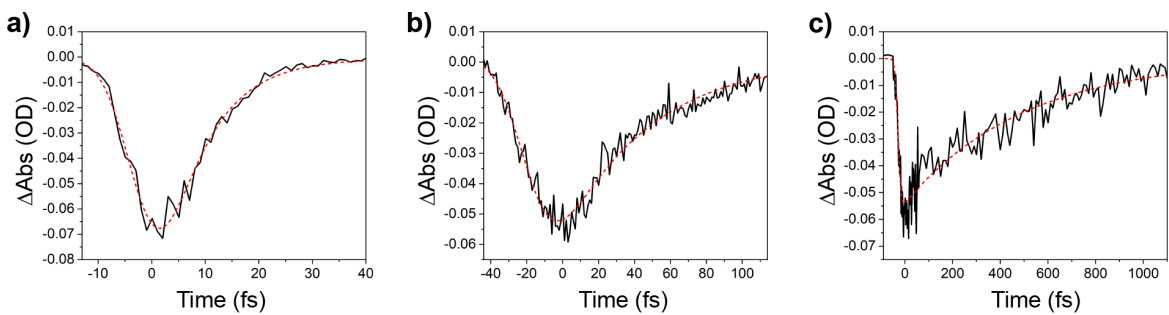

*Figure 4.* Time-resolved FWM signals of the (a) 3*s*, (b) 3*p*, and (c) 3*d* states are depicted with monoexponential fits to extract the measured autoionization decay times. The black lines represent the raw data and the red dashed lines represent the fitting functions. For this data set, the fitted decay results are 8 ± 2 fs for the 3*s* state, 50 ± 5 fs for the 3*p* state, and 503 ± 43 fs for the 3*d* state, with 2σ error reported.

The lifetimes of the 3*s*, 3*p*, and 3*d* states were measured by integrating the signals at ±1.6 mrad to obtain the transient intensities of each state. Example scans for each state are provided in Figure 4a, 4b, and 4c. The raw data (black solid line) was fitted to a convolution of an instrument response function with a monoexponential decay (red dashed line). While the results for these specific scans were 8 ± 2 fs for the 3*s* state, 50 ± 5 fs for the 3*p* state, and 503 ± 43 fs for the 3*d* state, the compiled and averaged autoionization decay lifetimes within this complete study were found to be 7 ± 2 fs, 48 ± 8 fs, and 427 ± 40 fs for the 3*s*, 3*p*, and 3*d* states, respectively, at 95% confidence intervals. These values were calculated from a collection of 4 scans for the 3*s* and 3*p* states, and 2 scans for the 3*d* state. Each reported lifetime is the average of the individual decay lifetimes provided by the fitting function, with the error reported as the square root of the sum of the variances for each individual scan, multiplied by 2 to obtain 95% confidence.

The experimentally measured and theoretically calculated autoionizing decay lifetimes for the 3*s* and 3*p* states are in good agreement with each other and with literature references, as shown in Table 1. For the 3*s* state, the NewStock, ASTRA, and referenced [16] lifetime results are all within the experimental error. For the 3*p* state, the ASTRA and all referenced lifetime [12–15] results fall within the experimental error here, however the 61 fs NewStock calculation lies 5 fs outside of the 95% experimental confidence interval. The experimentally retrieved lifetime of the 3*d* state is notably shorter than both theoretically calculated lifetimes.

One possibility to explain this discrepancy can be attributed to the limited range of motion (up to ~1 ps) of the piezoelectric delay stages used to control the NIR delays in the FWM apparatus, which did not allow data points to be taken at longer delays to capture the baseline after measuring the lifetime decay.

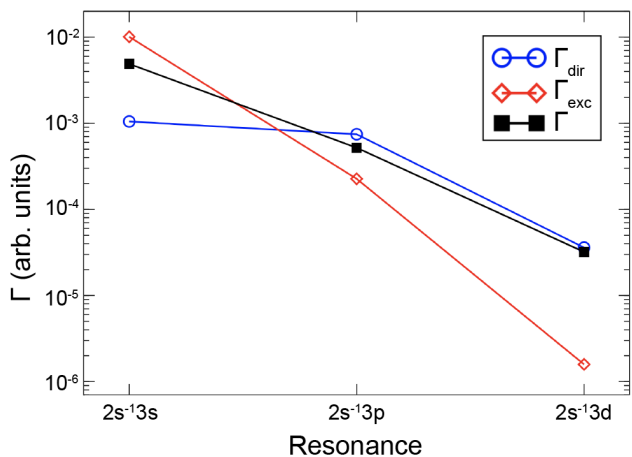

*Figure 5*. Hydrogenic model results for the $2s^{-1}3L$ autoionizing states width $\Gamma$ (black squares, total width). The contribution to the width only considering either the direct or the exchange term, $\Gamma_{dir}$ (blue circles) and $\Gamma_{exc}$ (red diamonds), respectively, is also shown.

To qualitatively understand the origin of the trend observed experimentally and reproduced by the *ab initio* calculations, we used a hydrogenic model to estimate the relative weight of the exchange ($\Gamma_{exc}$) and direct ($\Gamma_{dir}$) terms to the width of each resonance, which are obtained by switching off the contribution of the other term in Eq. [6]. Orbitals with angular momentum $l$ satisfy the boundary condition $r^l$ at the origin; this circumstance determines the penetration of the $3l$ orbitals in the internal portion of the ion, and hence their ability to exchange energy and momentum with core electrons. As shown in Figure 5, the exchange term dominates for the satellite electron in the 3s orbital, which has a finite value at the nucleus ($l = 0$), whereas the direct term dominates for the 3d orbital, which is vanishingly smaller in the nuclear region ($l = 2$). Since their direct and exchange decay amplitudes add coherently, however, due to their interference, both terms generally make a non-negligible contribution. Furthermore, the two terms can add destructively, with either of the two individual contributions being able to exceed the total. Both the direct and the exchange term exhibit a marked decrease as the orbital momentum of the electron satellite increases. This circumstance suggests that the short-range character of the interaction due to the Coulomb denominator is paramount, and hence that the trend in the resonance widths is to be attributed to how much the $n = 3$ orbitals penetrates the innermost shells.

*Table 1*. Summary of theoretical, experimental, and referenced autoionization decay lifetimes for neon's $2s2p^63s$, $2s2p^63p$, and $2s2p^63d$ states.

| Configuration | NewStock Theoretical Lifetime (fs) | ASTRA Theoretical Lifetime (fs) | Experimental Lifetime (fs) | References (fs) |
|---|---|---|---|---|
| $2s2p^63s$ | 5 | 6 | $7 \pm 2$ | 7 [16] |
| $2s2p^63p$ | 61 | 53 | $48 \pm 8$ | 50 [12], 41 [13], 55 [14], 47 [15] |
| $2s2p^63d$ | 571 | 679 | $427 \pm 40$ | n/a |

**Conclusions**

Attosecond noncollinear four wave mixing spectroscopy was used to measure the bright and dark state autoionization decay lifetimes of inner valence electronic excitations in neon atoms. Experimental lifetimes for the $2s2p^63s$, $2s2p^63p$, and $2s2p^63d$ states were measured to be $7 \pm 2$ fs, $48 \pm 8$ fs, and $427 \pm 40$ fs, respectively, with 95% confidence intervals. The results for the $3s$ and $3p$ states are in excellent agreement with the theoretical calculations and literature sources, while the experimental $3d$ state lifetime is somewhat shorter compared to the theoretical calculation due to the possibility of a systematic error because the time delay could not be taken out far enough. The overall trend of the lifetime results supports the predicted relationship that autoionization decay lifetime increases with less overlap in the radial region of the 2s core hole. This relationship was experimentally investigated here in the Ne $2s2p^63l$ states ($l=0$, 1, and 2) with few-femtosecond time resolution using attosecond FWM spectroscopy.

In conclusion, this work contains time-resolved dark state autoionization decay lifetime measurements using attosecond noncollinear FWM spectroscopy. Not only do the experiments establish the ability of FWM to investigate few-femtosecond autoionization dynamics, but it also benchmarks the viability of the FWM method to provide resonantly enhanced, time-resolved dynamics measurements in optically forbidden quantum states. Future applications of attosecond FWM spectroscopy will include studying core-excited states of more complex systems such as small molecules. As exemplified in this work, attosecond FWM provides access to optically forbidden states and therefore facilitates explorations of wider regions of electronic dynamics in atoms and molecules.

**Acknowledgements**

This work was performed by personnel and equipment supported by the Office of Science, Office of Basic Energy Sciences through the Atomic, Molecular and Optical Sciences Program of the Division of Chemical Sciences, Geosciences, and Biosciences of the U.S. Department of Energy at LBNL under Contract No. DE-AC02-05CH11231. Y-C.L. acknowledges financial support from the Taiwan Ministry of Education. J.G. acknowledges the Arnold and Mabel Beckman Foundation for support as an Arnold O. Beckman Postdoctoral Fellow. L.A., S.C., and C.M. acknowledge financial support from NSF theoretical AMO grant PHY-1912507 and from DOE CAREER grant No. DE-SC0020311. C.K. acknowledges support through the National Science Foundation Research Experiences for Undergraduates (REU) Grant Nos. EEC-1461231 and EEC-1852537.